# Performance analysis of Non Linear Filtering Algorithms for underwater images

*Dr.G.Padmavathi, Dr.P.Subashini, Mr.M.Muthu Kumar and Suresh Kumar Thakur*
*Department of Computer Science, Avinashilingam University for Women, Coimbatore, TN, India*

*Abstract* **Image filtering algorithms are applied on images to remove the different types of noise that are either present in the image during capturing or injected in to the image during transmission. Underwater images when captured usually have Gaussian noise, speckle noise and salt and pepper noise. In this work, five different image filtering algorithms are compared for the three different noise types. The performances of the filters are compared using the Peak Signal to Noise Ratio (PSNR) and Mean Square Error (MSE). The modified spatial median filter gives desirable results in terms of the above two parameters for the three different noise. Forty underwater images are taken for study.**

## I. INTRODUCTION

Generally underwater images are collected by image sensors. They are contaminated by the different type of noise. Basically three type of noise are present in underwater images namely, Speckle noise, impulse noise and Gaussian noise. There are many different cases of distortions. Underwater images suffer from quality degradation due to transmission of limited range of light, low contrast and blurred image due to quality of light and diminishing color. When an underwater image is captured, denoising is necessarily done to correct and adjust the image for further study and processing. Different filtering techniques are available in the literature for denoising of under water images. The performance of an image filtering system depends on its ability to detect the presence of noisy pixels in the image. Significant works have been done in both hardware and software to improve the signal-to-noise ratio for acoustic images.

In software, a denoising filter is used to remove noise from an image. Each pixel is represented by three scalar values representing the red, green, and blue chromatic intensities. When each pixel is studied, a smoothing filter takes into account the surrounding pixels to derive a more accurate version of this pixel. By taking neighboring pixels into consideration extreme "noisy" pixels can be replaced. However, outlier pixels may be uncorrupted with fine details, which may be lost due to the smoothing process. This paper evaluates five common smoothing algorithms for underwater images for three different noise types. This five smoothing algorithms are implemented in forty benchmark underwater images for different noise types. The simulation are done in matlab 7.1 version.

The organization of this paper is as follows: Section 2 is described a Methods of filter, Section 3 discusses about the Experiment as results and evaluation and Finally Section 4 gives conclusion.

## II. FILTERING METHODS

Conventionally linear filtering Algorithms were applied for image processing. The fundamental and the simplest of these algorithms is the Mean Filter as defined in (1). The Mean Filter is a linear filter which uses a mask over each pixel in the signal. Each of the components of the pixels which fall under the mask are averaged together to form a single pixel. This filter is also called as average filter. The Mean Filter is poor in edge preserving. The Mean filter is defined by

$$\text{Mean filter } (x_1 \ldots x_N) = \frac{1}{N}\sum_{i=1}^{N} x_i \quad \text{-----(1)}$$

where $(x_1 \ldots x_N)$ is the image pixel range. Generally linear filters are used for noise suppression. It gives minimum PSNR when compared to non linear filters. Which has to be generally maximum hence for underwater images non linear filters are taken for comparison. Median filter, Component Median filter, Vector Median filter, Spatial Median filter, Modified Spatial Median filters are compared for different type of noise.

*2.1 Median filter*

The Median Filter is performed by taking the magnitude of all of the vectors within a mask and sorted according to the magnitudes. The pixel with the median magnitude is then used to replace the pixel studied. The Simple Median Filter has an advantage over the Mean filter since median of the data is taken instead of the mean of an image. The pixel with the median magnitude is then used to replace the pixel studied. The median of a set is more robust with respect to the presence of noise. The median filter is given by

Median filter$(x_1....x_N)$ =Median $(\|x_1\|^2.......\|x_N\|^2)$ ------ (2)

When filtering using the Simple Median Filter, an original and the resulting filtered pixel of the sample have the same pixel. A pixel that does not change due to filtering is known as the root of the mask

**Advantage**
A major advantage of the median filter over linear filters is that the median filter can eliminate the effect of input noise values with extremely large magnitudes. (In contrast, linear filters are sensitive to this type of noise - that is, the output may be degraded severely by even by a small fraction of anomalous noise values).

2.2 Component Median Filter (CMF)
CMF is defined in (3), relies on the statistical median concept. In the Simple Median Filter, each point in the signal is converted to a single magnitude. In the Component Median Filter each scalar component is treated independently. A filter mask is placed over a point in the signal. When noise affects a point in a grayscale image, the result is called "salt and pepper" noise. In color images, this property of "salt and pepper" noise is typical of noise models where only one scalar value of a point is affected. For this noise model, the Component Median Filter is more accurate than the Simple Median Filter. The disadvantage of this filter is that it will create a new signal point that did not exist in the original signal, which may be undesirable in some applications. The CMF is defined by

$$CMF(x_1, ....x_N) = \begin{Bmatrix} .\text{Median } (X_{1r}..... X_{Nr}) \\ \text{Median } (X_{1g}..... X_{Ng}) \\ \text{Median } (X_{1b}...... X_{Nb}) \end{Bmatrix} \text{---- (3)}$$

**Advantage**
The Component Median Filter is more accurate than the Simple Median Filter.

2.3 The Vector Median Filter (VMF)

In the VMF, a filter mask is placed over a single point. The point with the minimum sum of vector differences is used to represent the point in the signal studied. The VMF is a well-researched filter and popular due to the extensive modifications. The VMF is defined by

VMF$(x_1,...x_N)$ =

$$\text{MIN}\left( \sum_{i=1}^{N} \| X_N - X_i \|,...., \sum_{i=1}^{N} \| X_N - X_i \| \right) \text{----- (4)}$$

**Advantage**
The advantage of this method is that filter outputs are close to each other and can be manipulated by 2D techniques.

2.4 Spatial Median Filter

The SMF is a new noise removal filter. The SMF and the VMF follow a similar algorithm and it will be shown that they produce comparable results.

The SMF is a uniform smoothing algorithm with the purpose of removing noise and fine points of image data while maintaining edges around larger shapes The spatial depth between a point and set of points is defined by,

$$S_{depth}(x, x_1... x_N) = 1\left(\left(\frac{1}{1-N}\right) \| \frac{\sum_{i-1}^{N} x - x_i}{x - x_i} \| \right) \text{------ (5)}$$

The SMF is an unbiased smoothing algorithm and will replace every point that is not the maximum spatial depth among its set of mask neighbors. The Modified Spatial Median Filter [1] attempts to address these concerns.

The following is the basic algorithm for determining the Spatial Median of a set of points, $x_1, ...,x_N$: Let $r_1, r_2, ..., r_N$ represent $x_1, x_2, ...,x_N$ in rank order such that it is used in basic algorithm for

$$\text{SMF } (x_1, ...... x_N) = r_1 \text{----------- (6)}$$

**Advantage**
The advantage of replacing every point achieves a uniform smoothing across the image. A good smoothing filter should simplify the image while retaining most of the original image shape and retain the edges.

*2.5 A Modified Spatial Median Filter for Gaussian noise [1]*

The SMF is similar to the VMF in that in both filters, the vectors are ranked by some criteria and the top ranking point is used to replace the center point. No consideration is made to determine if that center point is original data or not. The unfortunate drawback of these filters is the smoothing that occurs uniformly across the image. Across areas where there is no noise, uncorrupted data is removed unnecessarily.

In the Modified Spatial Median Filter (MSMF), after the spatial depth of each point within the mask is computed, an attempt is made to use this information to first decide if the mask's center point is an uncorrupted point. If the determination is made that a point is not corrupted, then the point will not be changed. The spatial depth of every point within the mask is calculated and then sorted based on depths in descending order.

By ranking these spatial depths in the set in descending order, a spatial order statistic of depth levels is created. The largest depth measures, which represent the collection of uncorrupted points, are pushed to the front of the ordered set. The smallest depth measures, representing points with the largest spatial difference among others in the mask

The MSMF is define by

$$\text{MSMF}(T, x_1, \ldots x_N) = \begin{cases} r_c & c = T \\ r_1 & c > T \end{cases} \quad \text{--------- (7)}$$

Where $r_c$ & $r_1$ are rank order

**Advantage**

Modified spatial Median filter median filter is the most suitable filter for denoising the images for different type of noise.

### III. EXPERIMENTAL SETUP AND EVALUATION

To test the accuracy of the non linear Filtering algorithms, three steps are followed.

i) First an uncorrupted underwater image is taken as input.
ii) Second different noises are added to the underwater image artificially.
iii) Third, the filtering algorithms are applied for reconstruction of underwater images.

To estimate the quality of reconstructed image, Mean Squared Error and Peak Signal to Noise Ratio are calculated for the original and the reconstructed images.

Performances of different filters are tested for three different types of noise models by calculating the Mean Square Error (MSE) and Peak Signal to Noise Ratio (PSNR). The values are calculated by the following expressions:

$$\text{PSNR} = 20 \log_{10} \frac{256}{sqrt(MSE)}$$

where MSE represents the mean square error of the estimation. The size of the image taken is 256X256 pixels.

The experiments are conducted using matlab 7.1 version for Gaussian noise, salt and pepper noise and speckle noise. The benchmark images used here are given in Annexure I.

- For each of forty images, different noise models are added artificially in the ratio 0.5.
- Next the different filtering algorithm is applied for each forty images with various noise models.
- The parameters PSNR and MSE value are calculated for the noise free image and noisy image.

Giving the reconstruction of an image, each image is assumed to have the dimensions of 256. The images in this contain a wide variety of subject matters and textures. Most of the images used are ship wreck, moor chain and mine in sonar images. The following fig (1a-1f) shows the average PSNR and MSE value for forty images.

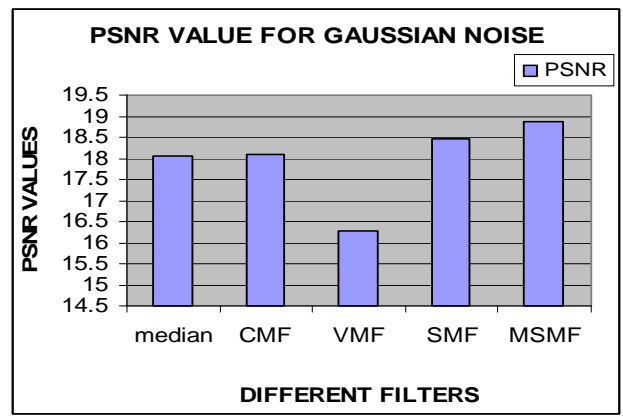

Fig -1a   PSNR for Gaussian noise

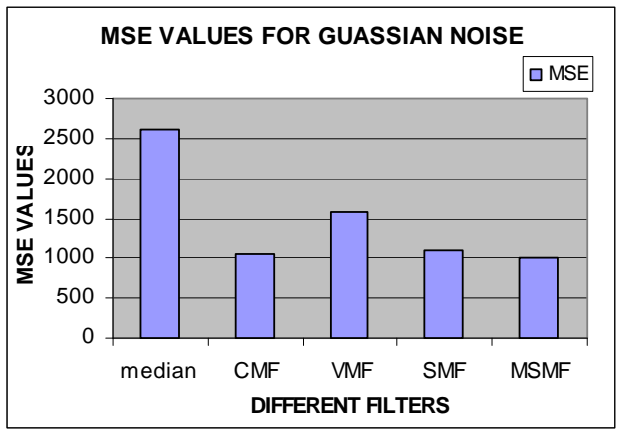

Fig -1b MSE for Gaussian noise

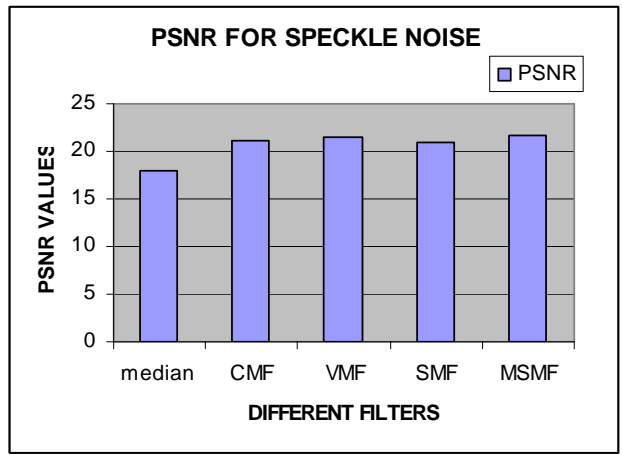

Fig 1-e PSNR for speckle noise

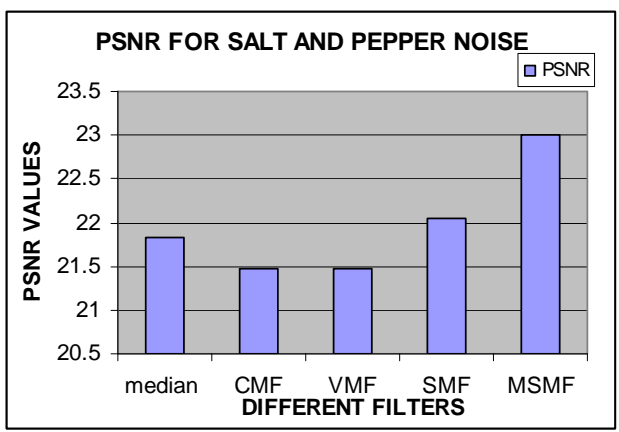

Fig -1c PSNR for salt and pepper noise

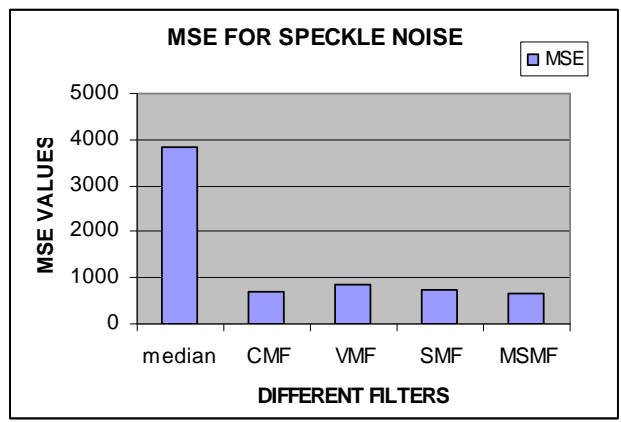

Fig -1f MSE for speckle noise
Fig – 1 Performance of different filters with various types of noise

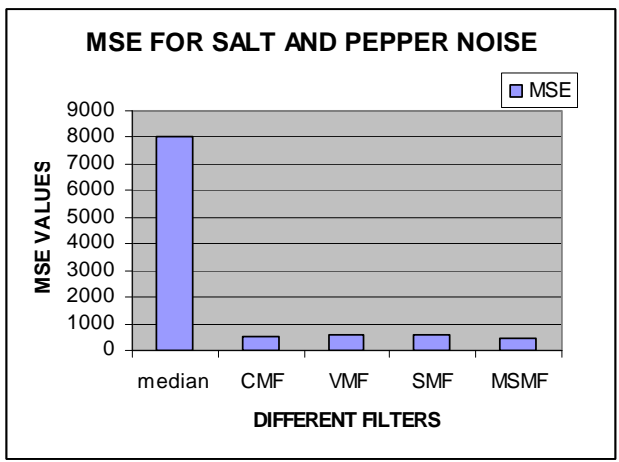

Fig -1d MSE for salt and pepper noise

Table .1 - comparison of filters using noise value and MSE

| Method | PSNR value | MSE value | Noise type |
|---|---|---|---|
| Median filter | 20.734458 | 3.83E+08 | Speckle noise |
| | 18.079603 | 2.61E+05 | Gaussian noise |
| | 21.849255 | 8.06E+07 | Salt & pepper noise |
| Component Median Filter(CMF) | 21.02803 | 683.67324 | Speckle noise |
| | 18.075605 | 1042.2113 | Gaussian noise |
| | 22.091855 | 532.57731 | Salt & pepper noise |
| Vector Median Filter(VMF) | 19.681275 | 8.51E+02 | Speckle noise |
| | 16.263548 | 1.59E+03 | Gaussian noise |
| | 21.478483 | 584.65543 | Salt & pepper noise |
| Spatial Median filter(SMF) | 20.994288 | 735.53547 | Speckle noise |
| | 18.474508 | 1108.2839 | Gaussian noise |
| | 22.047078 | 577.38177 | Salt & pepper noise |
| MSMF filter | 21.706158 | 656.5929 | Speckle noise |
| | 18.878773 | 998.99331 | Gaussian noise |
| | 23.01723 | 483.28943 | Salt & pepper noise |

The table.1 shows the comparison of five filters. The MSMF median type filters give better performance than other type of filters. It is clearly observed that salt & pepper noise is completely removed when compare to other two noise types. It is confirmed that Gaussian noise are removed poorly than other two noise types.

The test conducted on a dual core CPU with each processor running at 3GHz clock speed and 1GB of RAM. The program was run and compiled in matlab. The testing gives the result that MSMF performs better than other filters. The testing took roughly eight days to complete. The objective of test is to find a most siuted filter for underwater images.

## IV. CONCLUSION

In this paper image filtering algorithms are applied on images to remove the different types of noise that are either present in the image during capturing or injected in to the image during transmission. Underwater images when captured usually have Gaussian noise, speckle noise and salt and pepper noise. In this work, five different image filtering algorithms are compared for the three different noise types. The performances of the filters are compared using the Peak Signal to Noise Ratio (PSNR) and Mean Square Error (MSE). The modified spatial median filter (MSMF) gives desirable results in terms of the above two parameters for the three different noise types. Forty underwater images are taken for implementation.

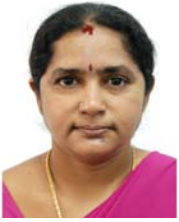
Dr. Padmavathi Ganapathi is the Professor and Head of Department of Computer Science, Avinashilingam University for Women, Coimbatore. She has 21 years of teaching experience and one year Industrial experience. Her areas of interest include Network security and Cryptography and real time communication. She has more than 72 publications at national and International level. She is a life member of many professional organizations like CSI, ISTE, AACE, WSEAS, ISCA, and UWA. She has five funded projects from UGC and DRDO

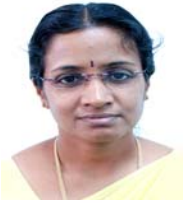
Dr. Subashini is the Lecturer (SG) in Department of Computer Science, Avinashilingam University for Women, Coimbatore. She has 15 years of teaching experience. Her areas of interest include Object oriented technology, Data mining, Image processing, Pattern recognition. She has 3 publications at national and International level.

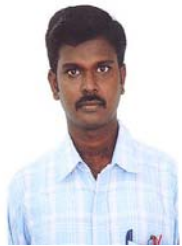
Mr.M.MuthuKumar. received the diplomo in ECE.from Arsan Ganesan Polytechnique College Sivaksi and B.E (EIE) degrees from KCET virudhunagar in 2004 and 2007 respectively.he is currently working as a Research staff in Department of Computer Science in Avinashilingam University for women and has one year of research experience. Her research interests are image and signal processing. he has 2 publications at national level.

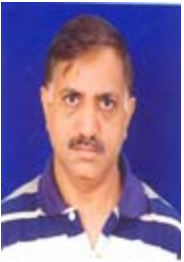
Mr.SK.Thakur is the Joint Director at Directorate of Naval Research & Development, DRDO HQ, New DelhiDeputy Director at Sectt. of Naval Research Board, New Delhi. He has More than 25 years of experience as a Naval Officer with Indian Navy. Experience in Policy Making & Goal Monitoring, Project Development & Monitoring, Equipment Maintenance & testing, Finance & Budgeting, Liaison with indigenous & foreign firms, Crisis & planned Management, Personnel & Administration; and Instructor/Teacher as Directing Staff; and Research & Development aspects.Experience of working with Research Personnel/Senior professors of IITs/IISc and Scientists from various R&D Institutions.and member of broad casting engineer society, new Delhi, member of IEEE and stragetic electronic group.

**Annexure –I**
**Underwater Images**

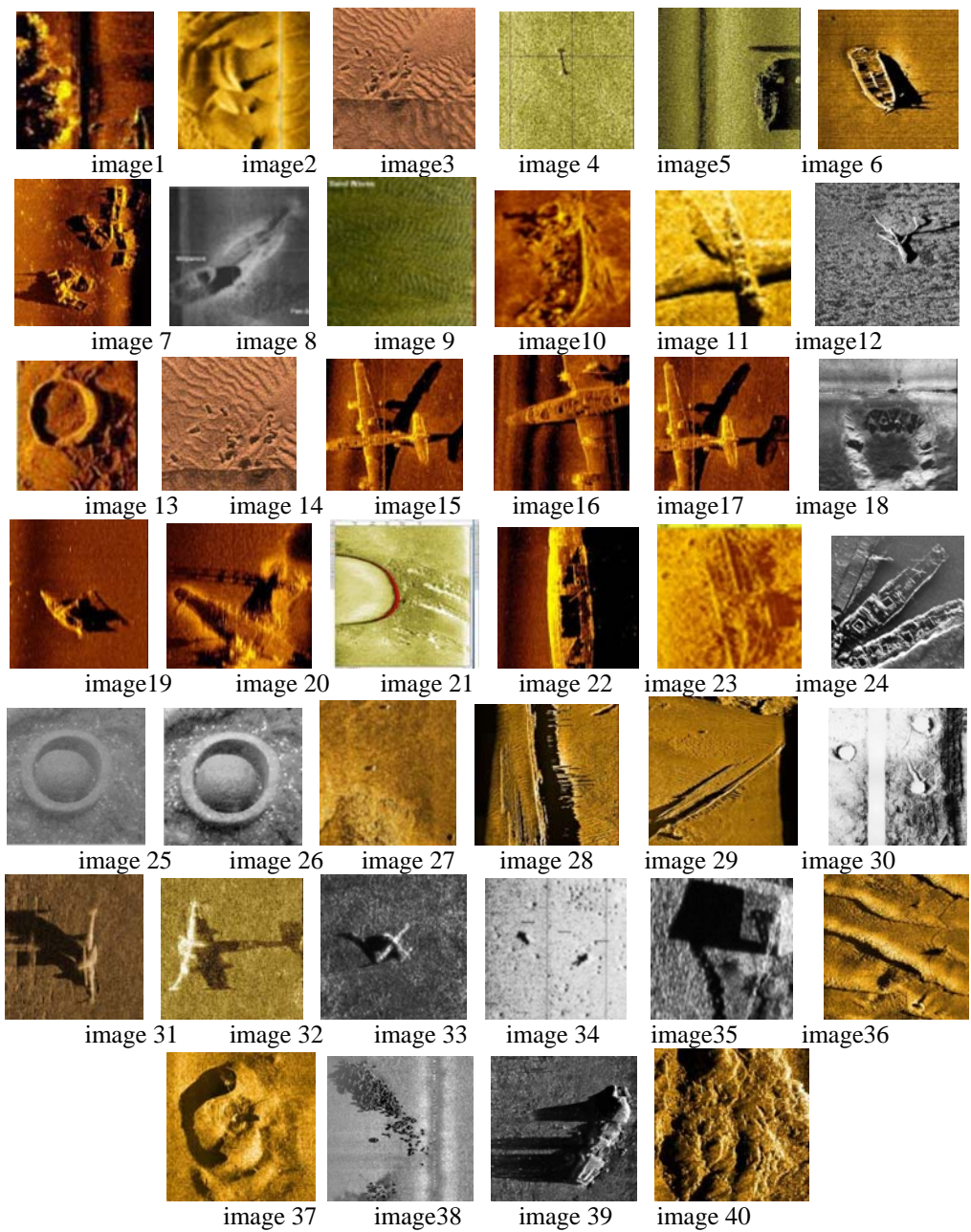